\documentclass[final]{ckm}                 

\confname{Workshop on the CKM Unitarity Triangle, IPPP Durham, April 2003}

\input mystyle.sum

\def\Eqn#1{Equation~(\ref{#1})}
\def\Fig#1{Figure~\ref{#1}}
\def\Tab#1{Table~\ref{#1}}

\newcommand{\Begeqn}{\begin{equation}}
\newcommand{\Endeqn}{  \end{equation}}

\newcommand{\Begeqnarray}{\begin{eqnarray}}
\newcommand{\Endeqnarray}{  \end{eqnarray}}

\newcommand{\Arraystretch}{1.2}

\newcommand{\Begtable}[3]{\begin{table}[#1]\caption{#2}\label{#3}
\begin{center}}
\newcommand{\Endtable}{\end{center}\end{table}}

\newcommand{\Begtabular}[1]{\renewcommand{\arraystretch}{\Arraystretch}
                         \begin{center}\begin{tabular}{#1}\hline\hline}
\newcommand{\Endtabular}{\hline\hline    \end{tabular}\end{center}}

\newlength{\Itemindent}
\def\Itemwidths{
\setlength{\leftmargini}{\Itemindent}
\setlength{\leftmarginii}{\Itemindent}
\setlength{\leftmarginiii}{\Itemindent}
\setlength{\leftmargin}{\Itemindent}
\settowidth{\labelwidth}{\Large$\bullet$}
\setlength{\labelsep}{\leftmargin}
\addtolength{\labelsep}{-\labelwidth}}

\def\Itemspace{
  \topsep    0pt plus 1pt minus 1pt             
  \partopsep 0pt plus 1pt minus 1pt             
  \parskip   0pt plus 1pt minus 1pt 
  \parsep    0pt plus 1pt minus 1pt 
  \itemsep   1pt plus 1pt minus 1pt}

\catcode`\@=11 

\newcommand{\Begitem}{\Itemwidths \begin{list}
{\csname\@itemitem\endcsname}
{\ifnum \@itemdepth >4 \@toodeep\else \advance\@itemdepth 1 
 \edef\@itemitem{labelitem\romannumeral\the\@itemdepth}
 \Itemspace \fi}}
\newcommand{\Enditem}{\end{list}}

\catcode`\@=12 

\def\Cleoiii{CLEO~III}
\def\Cleoc{CLEO-c}
\def\Cesrc{CESR-c}

\def\Vud{|V_{ud}|}

\def\Vub{|V_{ub}|}
\def\Vcd{|V_{cd}|}
\def\Vcs{|V_{cs}|}
\def\Vcb{|V_{cb}|}
\def\Vtd{|V_{td}|}

\def\Vtb{|V_{tb}|}

\def\Vcq{|V_{cq}|}

\def\fB{f_B}
\def\fDq{f_{\Dq}}

\def\calB{{\cal B}}
\def\calE{{\cal E}}
\def\calF{{\cal F}}
\def\calG{{\cal G}}

\def\calO{{\cal O}}

\def\Lunits{cm$^{-2}$ s$^{-1}$}

\def\fbinv{fb$^{-1}$}

\def\nubar{\bar{\nu}}

\def\nue{\nu_e}
\def\numu{\nu_\mu}

\def\ep{e^+}

\def\mup{\mu^+}

\def\ellp{\ell^+}
\def\ellm{\ell^-}

\def\mup{\mu^+}

\def\taup{\tau^+}

\def\pip{\pi^+}
\def\pim{\pi^-}
\def\piz{\pi^0}

\def\Km{K^-}

\def\Kzbar{\bar{K}^0}

\def\Dp{D^+}
\def\Dm{D^-}
\def\Dz{D^0}
\def\Dbar{\bar{D}}
\def\Dzbar{\bar{D}^0}
\def\Ds{D_s}
\def\Dsbar{\bar{D}_s}
\def\Dsp{D_s^+}
\def\Dsm{D_s^-}
\def\Dqp{D_q^+}
\def\Dq{D_q}

\def\Dstar{D^*}

\def\DztoKmpip{\Dz \to \Km\pip}
\def\DptoKmpippip{\Dp \to \Km\pip\pip}

\def\Jpsi{J/\psi}

\def\Bp{B^+}

\def\Bz{B^0}
\def\Bbar{\bar{B}}
\def\Bzbar{\bar{B}^0}

\def\BtoXsgamma{B \to X_s \gamma}

\def\BtoXlnu{\Bbar \to X \ell \nubar}
\def\BtoXclnu{\Bbar \to X_c \ell \nubar}
\def\BtoXulnu{\Bbar \to X_u \ell \nubar}

\def\BtoDstarlnu{\Bbar \to \Dstar \ellm \nubar}

\def\Jpsi{J/\psi}

\def\fbinv{fb$^{-1}$}

\def\Jpsi{J/\psi}

\def\fDp{f_{\Dp}}
\def\fDs{f_{\Ds}}

\def\FDstar{\calF_{\Dstar}}

\def\eg{{\it e.g.}}

\def\etal{{\it et al.}}

\def\ltsim{\raisebox{-.55ex}{\rlap{$\sim$}} \raisebox{.45ex}{$<$}}
\def\gtsim{\raisebox{-.55ex}{\rlap{$\sim$}} \raisebox{.45ex}{$>$}}

\title{Contributions of Charm Physics to CKM Parameters}

\author{D.G. Cassel}
\address{Laboratory for Elementary-Particle Physics, Cornell University,
Ithaca, NY 14853}

\begin{document}

\begin{abstract}Determinations of CKM matrix elements are clouded by uncertainties in
nonperturbative QCD parameters that relate measurable quantities to the underlying
parton-level processes.  A principal goal of the \Cleoc\ program is to provide precision
measurements in the $c$-quark sector that will stimulate lattice QCD theorists
to calculate relevant nonperturbative QCD parameters in this sector and to validate the
calculations.  This interaction between theory and experiment should build confidence in
calculations of the parameters in the $b$-quark sector required for precision
determinations of the CKM matrix elements $\Vcb$, $\Vub$, and $\Vtd$. 
\end{abstract}

\maketitle

\section{Introduction and Motivation}

Everyone at this workshop is all too aware that at least one nonperturbative QCD
parameter that relates measurable quantities to the underlying parton-level
processes stands between an experimental measurement and determination of a CKM matrix
element.  Progress in determining CKM matrix elements is already -- or very soon will
be -- limited by uncertainties in these QCD parameters.  Theoretical uncertainties
totally dominate the $\Vtd$ error and -- even with the relatively modest CLEO 
luminosities  -- they are significant in $\Vcb$ and $\Vub$ errors.  Experimental
uncertainties will decrease significantly when the enormous BaBar and Belle data
samples are fully understood and evaluated.  Full exploitation of these data samples
for determining CKM matrix elements will require substantial theoretical progress in
developing reliable methods for calculating these nonperturbative parameters.

Lattice QCD (LQCD) shows promise of being a theory capable of calculating most of the
required parameters to a precision of a few percent \cite{hplqcd}.  However,
verification of these calculations will require comparison of LQCD results with a large
number and wide variety of precision experimental measurements in the $c$- and 
$b$-quark sectors.  Providing precise $c$-quark decay data to motivate and validate
theoretical progress in nonperturbative heavy quark physics is a major focus of the
\Cleoc\ program
\cite{cleocyb}.  The other major focus -- searches for glueballs -- is not directly
related to the subject of this workshop.

\section{Determining $\Vtd$ from $B^0 \bar{B}^0$ Mixing}

Determination of $\Vtd$ from $B^0 \bar{B}^0$ mixing is the extreme example of the
mismatch between experimental and theoretical precision \cite{ckmyb}.  The measured mass
difference due to $\Bz\Bzbar$ mixing is related to  $\Vtd$ by\\[-3ex]
\Begeqn
\Delta m_d
= {G_F^2 \over 6 \pi^2}\, \eta_{QCD}\, M_B
{f^2_B B_B}\, m_t^2\, F(x_t)\, \Vtd^2\, \Vtb^2 \label{eq:deltamd}
\Endeqn
where
$\Delta m_d$ is  the $\Bz$ mass difference,
$G_F$ is         the Fermi constant,
$\eta_{QCD}$ is  a QCD correction factor,
$M_B$ is         the $B^0$ mass,
$B_B$ is         the $B^0$ bag constant,
$f_B$ is         the $B^0$ pseudoscalar decay constant,
$m_t$ is         the top-quark mass, and
$F(x_t)$ is      a known function of $x_t = m_t^2/m_W^2$.
Everything in this expression is reasonably well known, except $\Vtd$, 
$f_B$, and $B_B$. \Tab{tab:sigmaVtd} gives the principal contributions to the
uncertainty in $\Vtd$ using parameters from the most recent Particle Data Group CKM
review \cite{pdg}.  The contribution from the theoretical uncertainty in
$\sqrt{B_B} f_B$ dominates the contribution from the $\Delta m_d$ error by an order of
magnitude and the contribution from the $m_t$ error by nearly an order of magnitude! 
\vspace*{-4ex}
\Begtable{!htb}{The principal sources of uncertainties in $\Vtd$ and their
contributions to $\Delta \Vtd$.  The values of $\Vtd$ and $\Delta \Vtd$ (in
the lazy-L-shaped region of the table) have been multiplied by
$10^3$.}{tab:sigmaVtd}
\Begtabular{lcccc}
 & $\Delta m_d$ & $m_t$ & $\sqrt{B_B} f_B$ & \multicolumn{1}{|c}{$\Vtd$}
\\[-1ex]
 &   [ps$^{-1}$]   & [Gev] &    [MeV]   & \multicolumn{1}{|c}{ }   \\ 
Value & $0.489 \pm 0.008$ & $166 \pm 5$ & {$226 \pm 36$} & 
\multicolumn{1}{|c}{8.44} \\ 
\cline{1-4}
\\[-2.5ex]
$\Delta \Vtd$ &
               $^{+0.06}_{-0.06}$ &
               $^{+0.18}_{-0.18}$ & {$^{+1.5}_{-1.1}$}  &
$^{+2.0}_{-1.4}$\\[1ex] 
\Endtabular
\Endtable
\vspace*{-7ex}

\subsection{Determining $\fB$}

\begin{figure}[htb]
\hbox to\hsize{\hss
\includegraphics[width=0.82\hsize]{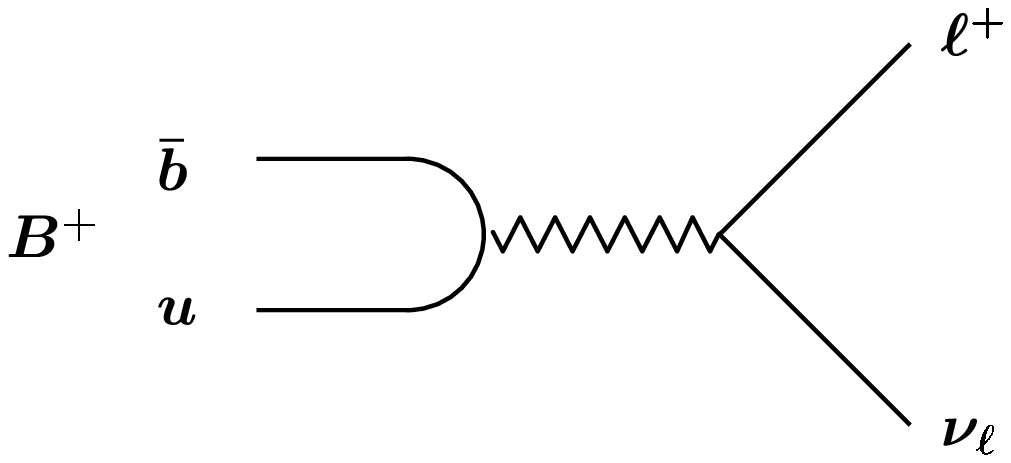}
\hss}
\caption{The Feynman diagram for $\Bp \to \ellp\nu$ decay.}
\label{fig:Bptolnu}
\end{figure}

The factor $f_B\Vub$ occurs in the decay amplitude for the $\bar{b}uW$ vertex in
leptonic $B$ decay, illustrated in \Fig{fig:Bptolnu}.  The decay width for leptonic
$B^+$ decays is 
\Begeqn
 \Gamma(\Bp \to \ell^+ \nu_\ell) = {G_F^2 \over 8\pi}\,
M_B\, m_{\ell}^2 \left(1 - {m_{\ell}^2 \over M^2_B} \right)
{\fB ^2\, \Vub^2}  \label{eq:fb}
\Endeqn
where $M_B$ is the $\Bp$ mass and $m_\ell$ is the $\ellp$ mass.  Hence, 
measurement of $\calB(\Bp \to \ell^+ \nu_\ell)$ would determine 
$\fB  \Vub$.   However, there are serious problems with determining $f_B$ this
way.  First, because $\Vub$ is very small, the leptonic branching fractions are also very
small for a reasonable value (200 MeV) of $\fB$:\\[1ex]
$\calB(\Bp \to \mup\nu_\mu) \sim 3 \times 10^{-7}$\\[1ex]
$\calB(\Bp \to \taup\nu_\tau) \sim 6 \times 10^{-5}$.\\
Second, detection of these decays requires reconstruction of the neutrino(s)
using the complete reconstruction of a tagging hadronic $B$ decay.  There are very
many $B$ decay modes with small branching fractions and small reconstruction
efficiencies due to the necessity of reconstructing the $D$ daughters from
the $B$ decays.  Hence, in the foreseeable future, LQCD will certainly
be required for precision estimates of $f_B$.  

\subsection{Determining $\Vcb$ from Exclusive\\ Semileptonic $B$ Decay}

Inclusive and exclusive $\BtoXlnu$ decay can be used to determine $\Vcb$ and $\Vub$
\cite{ckmyb}.  The status of inclusive measurements of $\Vcb$ -- based on CLEO's recent
measurements of some nonperturbative parameters using moments of $\BtoXsgamma$ and
$\BtoXclnu$ -- are described elsewhere in this workshop \cite{dgcmoments}.
Measurements of $\Vub$ are described in the 2002 CKM workshop \cite{ckmyb}
and in other reports in this workshop \cite{lkg}.
Hence, in this report I will concentrate on exclusive measurements of $\Vcb$ and
mention corresponding measurements of $\Vub$.

\begin{figure}[htb]
\hbox to\hsize{\hss
\includegraphics[width=0.82\hsize]{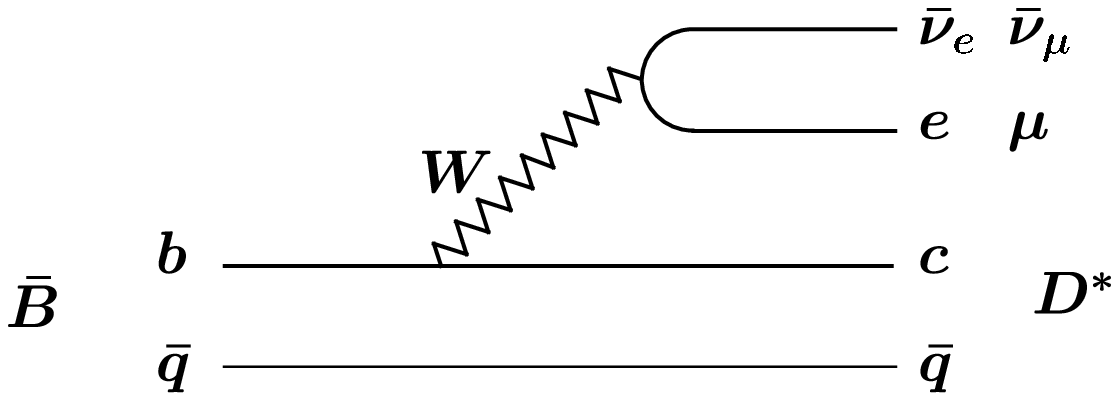}
\hss}
\caption{The Feynman diagram for $\BtoDstarlnu$ decay.}
\label{fig:BtoDstarlnu}
\end{figure}\vspace*{-1ex}

The Feynman diagram for $\BtoDstarlnu$ decay is illustrated in \Fig{fig:BtoDstarlnu}.
From Heavy Quark Effective Theory (HQET) and Isgur-Wise Symmetry, the differential
decay width for $\BtoDstarlnu$ decay is
\Begeqn
{d \Gamma(w) \over dw} = 
{G_F^2 \over 48 \pi^3}\; \calG(w)\;  \Vcb^2~
{ \FDstar^2(w)}\label{eq:BDstarlnu}
\Endeqn
\vspace*{-3ex}
\[ \hspace*{-\mathindent}{\rm with}\; w \equiv v_B \cdot v_{\Dstar} = 
{\calE_{\Dstar}
\over M_{\Dstar}}  = { M_B^2 + M_{\Dstar}^2 -q^2 \over 2 M_B M_{\Dstar}}. \]
\vspace*{-2ex}
In these expressions, $M_B$ and $M_{\Dstar}$ are the masses of the $B$ and $\Dstar$,
$v_B$ and $v_{\Dstar}$ are the four-velocities of the $B$ and $\Dstar$,
$\calE_{\Dstar}$ is the energy of the $\Dstar$ in the $B$ rest frame, 
$\calG(w)$ is a known function of $w$, $\FDstar(w)$ is an unknown form factor, and
$q^2$ is the square of the invariant mass of the $W$ or the $\ell\nu$ system. 

Since everything else in \Eqn{eq:BDstarlnu} is known or can be measured, the product
$\Vcb\FDstar(w)$ can be measured.  In particular -- with sufficient data -- the $w$
dependence of $\FDstar(w)$ can be determined accurately.   However, to determine $\Vcb$
we still need $\FDstar(1)$ from theory.  This quantity is constrained by HQET,
$\FDstar(1) \approx \eta_A [ 1 + \calO(1/m_Q^2) ]$, for large heavy quark masses and can
be computed with LQCD.  However, even with current experimental
uncertainties \cite{ckmyb,bdstarlnu}, the uncertainty in this parameter makes a
significant contribution to the uncertainty in determining $\Vcb$ by this method.

Determining $\Vub$ from exclusive $\BtoXulnu$ decays is even worse.  We 
don't even have HQET and Isgur-Wise Symmetry to constrain the form factors and 
theoretical uncertainties dominate current measurements \cite{bulnu}.  Hence, to
measure $\Vub$, we need a reliable theory for the form factors $f(q^2)$
for these decays.

\section{The \Cesrc/\Cleoc\ Program} 

\Cleoc\ is a focused program of measurements and searches in $e^+e^-$ collisions in
the $\sqrt{s}=3-5$ GeV energy region.  The items in the program most relevant
for this workshop are measurements of: absolute charm branching fractions, 
$D$ meson semileptonic decay form factors, $\Vcd$ and $\Vcs$, and the  decay
constants $f_D$ and $f_{D_s}$.  Other items in the core \Cleoc\ program
include: searches for new physics, \eg, $CP$ violation in $D$ decay, rare $D$
decays, and $D\bar{D}$ mixing without backgrounds from doubly suppressed Cabibbo
decays; and QCD studies, particularly $b\bar{b}$ spectroscopy and searches for
glue-rich exotic states (glueballs) \cite{cleocyb}.

The existing state-of-the-art \Cleoiii\ detector is a crucial element of this program
\cite{cleocyb}. It includes a central drift chamber for measuring the momenta of
charged particles and identifying them via their energy losses ($dE/dx$), a CsI
electromagnetic calorimeter for photon detection and electron identification, and a
ring imaging Cherenkov detector (RICH) for charged particle identification at higher
momenta.  The capabilities and performance of this detector represent substantial
advances above those of other detectors that have operated in the charm threshold
region.

\subsection{\Cleoc\ Run Plan}

The core \Cleoc\ program consists of four components, each expected to take
about one year to complete.  The anticipated data samples are:
\Begitem
\item Prologue -- $\Upsilon(nS)$'s $\gtsim 1.2\,$ \fbinv\ each
\Begitem
\item $\Upsilon(1S)$, $\Upsilon(2S)$, $\Upsilon(3S)$ -- {Completed}
\item 10-20 $\times$ the previous world's data 
\Enditem
\item Act I-- $\psi(3770)$ 3 \fbinv\
\Begitem
\item 30 M $D\Dbar$ events, 6 M tagged $D$'s
\item 310$\times$MARK III data
\Enditem
\item Act II -- $\sqrt{s} \sim 4.1$ GeV -- 3 \fbinv
\Begitem
\item 1.5 M $\Ds\Dsbar$ events, 0.3 M tagged $\Ds$'s
\item 480$\times$MARK III data and 130$\times$BES II data
\Enditem
\item Act III -- $\Jpsi$ -- 1 \fbinv\
\Begitem
\item 1 G $\Jpsi$ decays
\item 170$\times$MARK III data and 20$\times$BES II data
\Enditem
\Enditem
Taking data at the narrow $\Upsilon$ resonances is complete.  
Goals of this program include: precision measurement of matrix elements, $\Gamma$, and
$\Gamma_{ee}$ to compare with LQCD calculations; and $b\bar{b}$ spectroscopy
studies including searches for $\eta_b$, $h_b$, and $\Upsilon(1D)$ states.  The
$\Upsilon(1D)$ has already been observed in these data \cite{ichep:tomasz}.

At the time of this workshop we are in the midst of a shutdown to replace the \Cleoiii\
silicon vertex detector with a low-mass gaseous vertex detector and to upgrade CESR
for high luminosity in the charm threshold region.

\subsection{The \Cesrc\ Upgrade}

Running at all energies from the $\Jpsi$ to above the $\Upsilon(4S)$ is possible
with  existing superconducting interaction region quadrupole magnets.  We
have already taken modest amounts of data -- comparable to some previous data samples --
at the  $\psi(2S)$ and $\psi(3770)$. 
 
In the $\Upsilon$ region, synchrotron radiation damping reduces the size of beams in
CESR and is a crucial factor for achieving high luminosity.  This damping will be
substantially reduced at the lower energies in the charm threshold region, resulting
in serious reduction of luminosity.  Much of this luminosity loss can be recovered by
installing wiggler magnets (magnets with alternating magnetic field directions) to
increase synchrotron radiation.  We will use superferric wiggler magnets (Fe poles and
superconducting coils). We require a total of 12 of
these magnets, each 1.7 m long with 8 poles and maximum field  2.1~T.   We designed
and built a prototype superferric wiggler and installed\break\newpage
 it in CESR.  In
fact we took our low energy $\psi(2S)$ and $\psi(3770)$ data using this wiggler.  The
wiggler performed as expected and gave us the confidence we needed to proceed with the
CESR upgrade.  (Not entirely coincidentally, these magnets are excellent prototypes
for the damping ring wigglers required in a future linear collider.)  The first 6
wigglers will be installed by the end of the current shutdown.  These wiggler magnets
are the most substantial hardware upgrade in the
\Cleoc/\Cesrc\ program.
  
The luminosity we can achieve in the charm threshold region will still be below that
achieved in the $\Upsilon$ region ($\gtsim 1\times 10^{33}$~\Lunits).  We
anticipate luminosities ranging from $0.2\times 10^{33}$~\Lunits\ at 3.1 GeV to
$0.4\times 10^{33}$ ~\Lunits\ at 4.1 GeV. 

\section{Studying the \Cleoc\ Physics Reach}

Using a fast parameterized Monte Carlo program, we studied the ability of the
\Cleoc\ program to address many of the most important physics questions whose
answers may lie in the charm threshold region.  The parameters of the program were
carefully tuned to match the achieved performance of the \Cleoiii\ detector.  In the
following sections, I summarize a few of the conclusions of these studies.  These
studies, the performance of the \Cleoc\ detector, and the CESR upgrade plans are
described in much more detail  -- with comprehensive references -- in the
\hbox{\Cleoc}/\Cesrc\ project description \cite{cleocyb}.

\section{Hadronic $D$ Decays in \Cleoc}

\begin{figure*}[!tb]
\begin{center}
\includegraphics[width=0.45\hsize]{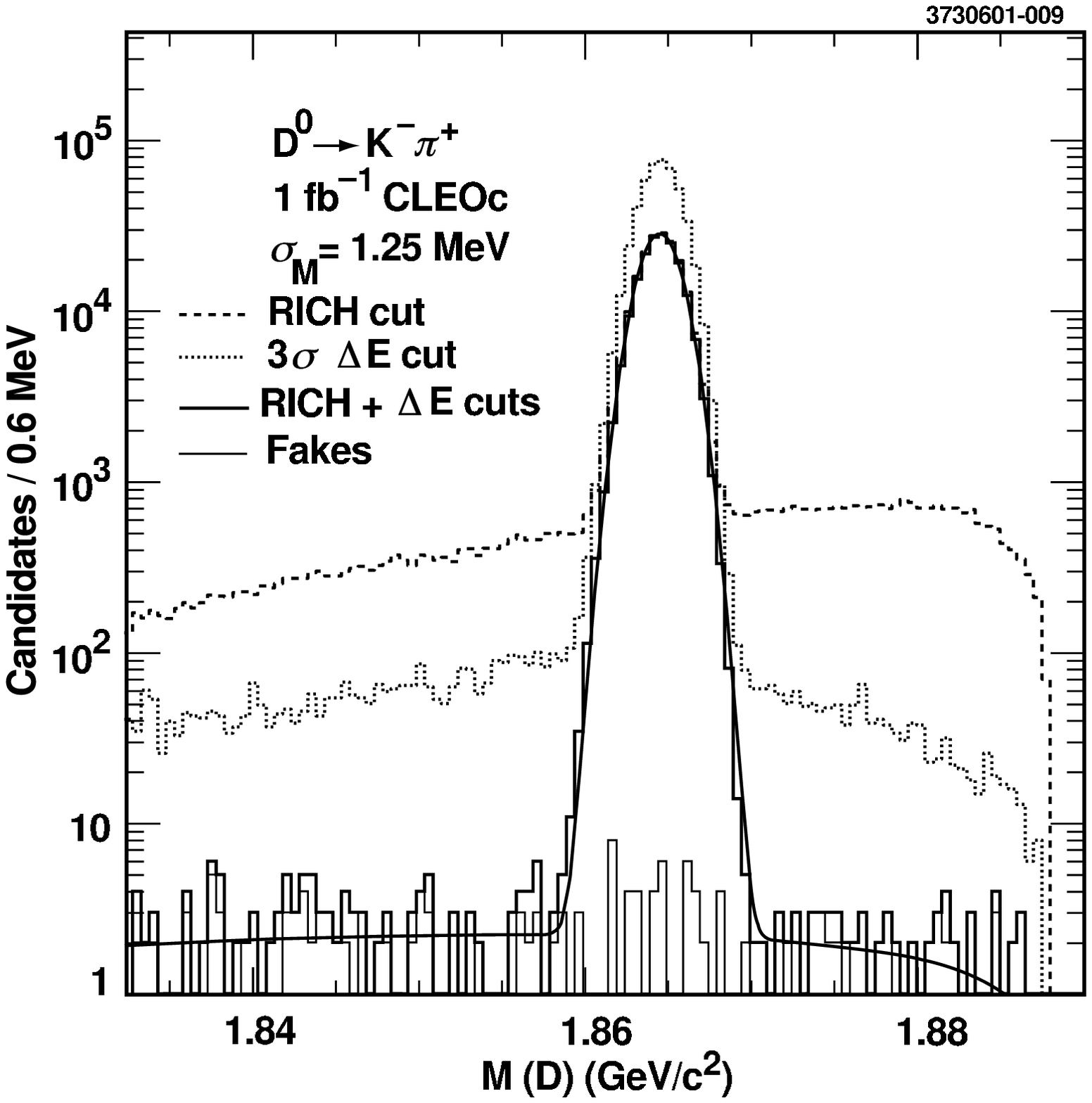}
\hfill\includegraphics[width=0.45\hsize]{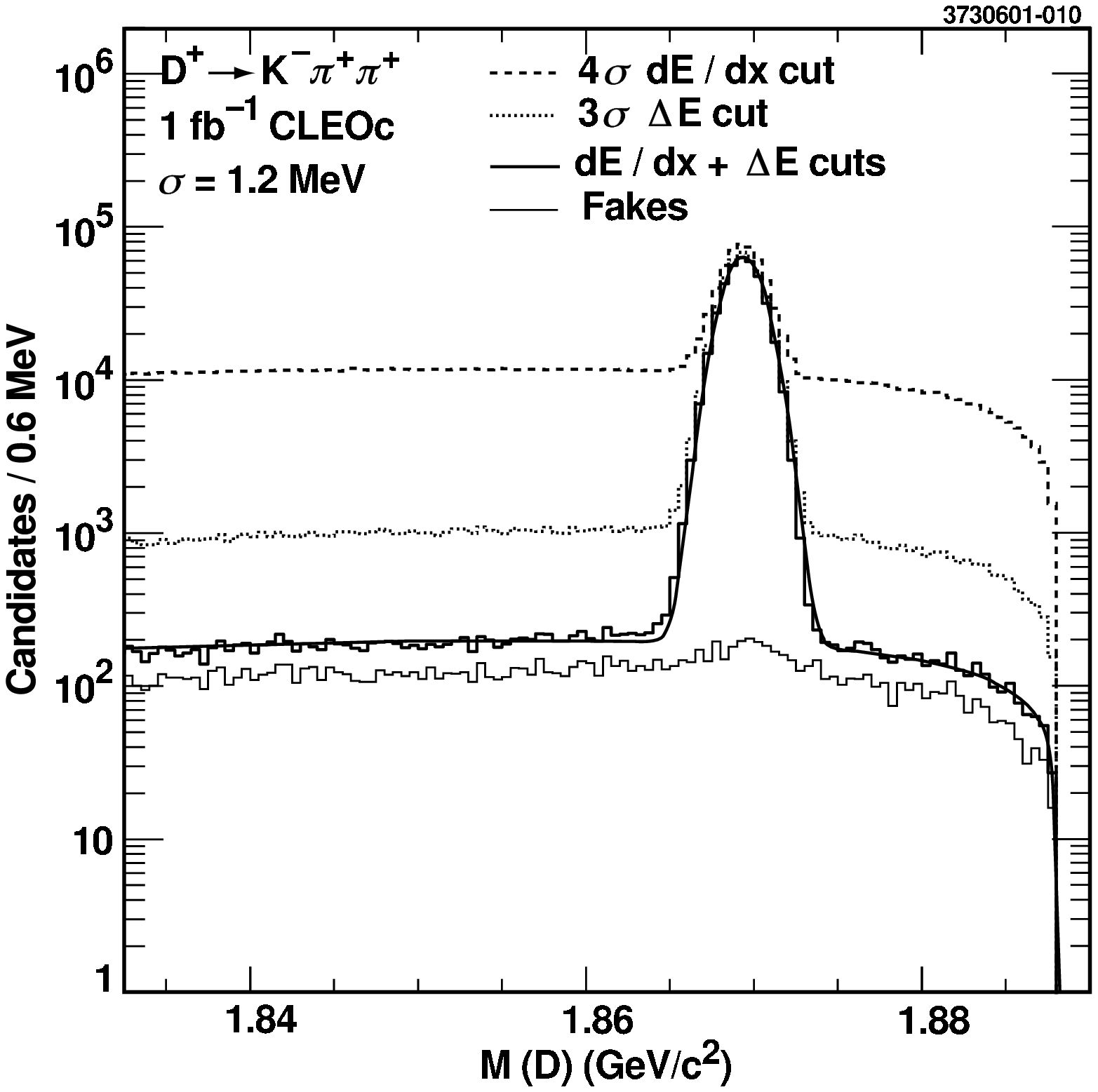}
\end{center}
\caption{Reconstructed $D$ mass distributions for (left) $\DztoKmpip$ and (right)
$\DptoKmpippip$ decays from Monte Carlo simulations.  Note the logarithmic scales, low
backgrounds, and that the Monte Carlo samples correspond to only 1 \fbinv\ of data,
instead of the anticipated 3 \fbinv.}
\label{fig:drecon}
\end{figure*}

\def\percenticks{									
	\majortick{	0	}{	0	}		\minortick{	33	}
	\majortick{	67	}{	5	}		\minortick{	100	}
	\majortick{	133	}{	10	}		\minortick{	167	}
	\majortick{	200	}{	15	}		\minortick{	233	}
	\majortick{	267	}{	20	}		\minortick{	300	}
	\majortick{	333	}{	25	}		\minortick{	367	}
	\majortick{	400	}{	30	} 
 }

\begin{figure*}[bt!]
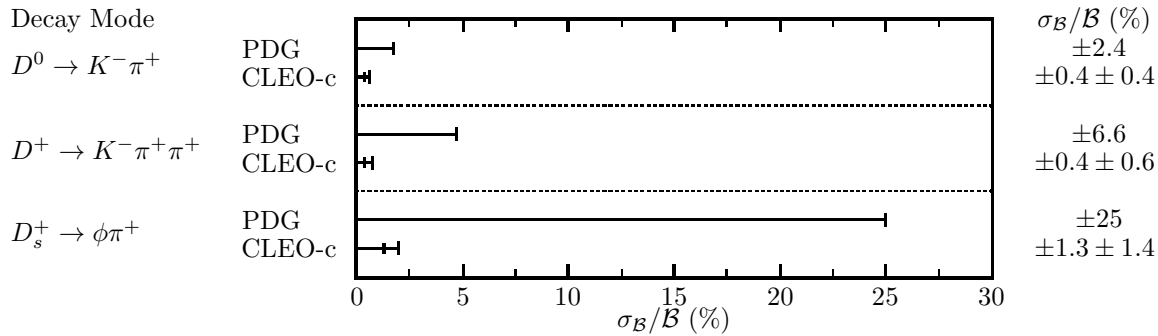

\begin{center}
\setlength{\unitlength}{0.6pt}
\newsummarypar{		400	}{205}{105}{0}{0}{18}																				
\begsummary																					
\titline{Decay Mode}{$\sigma_{\cal B}/{\cal B}$ (\%)}																							
\datline{\hfill	 PDG~~~		}{ $		 \pm 			2.4	$} \onlyerrbar{	0	}{	0	}{	23	} 
\datback
\datline{$D^0 \to K^- \pi^+$}{ }
\datback				
\datline{\hfill	 CLEO-c}{$		 \pm 	0.4	 \pm 	0.4	$} \onlyerrbar{	0	}{	5	}{	8	} 
\divdash																							
\datline{\hfill	 PDG~~~		}{ $		 \pm 			6.6	$} \onlyerrbar{	0	}{	0	}{	63	} 
\datback
\datline{$D^+ \to K^- \pi^+ \pi^+$}{ }
\datback				
\datline{\hfill	 CLEO-c}{$		 \pm 	0.4	 \pm 	0.6	$} \onlyerrbar{	0	}{	5	}{	10	} 
\divdash																							
\datline{\hfill	 PDG~~~		}{ $		 \pm 			25	$} \onlyerrbar{	0	}{	0	}{	333	} 
\datback			
\datline{$D_s^+ \to \phi \pi^+$}{ }
\datback																							
\datline{\hfill	 CLEO-c}{$		 \pm 	1.3	\pm	1.4	$} \onlyerrbar{	0	}{	17	}{	26	} 
\drawframe{$\sigma_{\cal B}/{\cal B}$ (\%)}																							
\percenticks																							
\endsummary																							
\end{center}\vspace*{-1ex}
\caption{Relative errors in the $D$ meson reference branching fractions. 
PDG errors are from PDG 2003 and CLEO-c errors are errors expected from CLEO-c.}
\label{fig:dbferr}
\end{figure*}\vspace*{-1ex}

Reconstructing exclusive hadronic decays of $D$ mesons is the foundation of the
\Cleoc\ charm physics program.  Hadronic decay modes can be reconstructed very cleanly
in the \Cleoc\ detector as illustrated in \Fig{fig:drecon}.  Although these modes are
the simplest $\Dz$ and $\Dp$ decay modes to reconstruct, we studied much more
complicated modes and found that we will also be able to reconstruct many
higher-multiplicity modes with very small backgrounds.  These exclusive hadronic decays
can then be used to tag $D\bar{D}$ events and provide clean samples of $D$ or $\bar{D}$
decays for measuring hadronic decay branching fractions or studying semileptonic and
leptonic $D$ decays.  

Absolute $D$ branching fractions can be measured by comparing double tag
($D\Dbar$) rates to  single tag ($D$ or $\Dbar$) rates -- a technique pioneered by
MARK~III \cite{markiii}.  Most systematic errors cancel with this technique and
knowledge of production rates is not required.  In our Monte Carlo studies we find
that double tag events are very clean with little background.

\newpage

\Begtable{!hbt}{Anticipated single and double tag rates for reference $D$ meson
branching fractions and the expected statistical and total errors on the
branching fractions. These rates are for the full 3 \fbinv\
\Cleoc\ data samples.}{tab:dtags}
\Begtabular{lcccc}
~~~~~  & ~Single~ & ~Double~ & ~Statistical~ & ~Total~ \\
  & Tags   &  Tags   & Error      & Error \\                   
\hline
$\Dz$   & 0.54 M & 53,000 & 0.4\% &  0.6\% \\
$\Dp$   & 1.14 M & 60,000 & 0.4\% &  0.7\% \\
$\Dsp$  & 0.15 M & ~6,000 & 1.3\% &  1.9\% \\
\Endtabular
\Endtable
\vspace*{-3ex}

The hadronic branching fractions, $\calB(\Dz \to \Km\pip)$, 
$\calB(\Dp \to \Km \pip\pip)$, and $\calB(D^+_s \to \phi\pi^+)$ are the
ref-\break\newpage
erence branching fractions for all $D$ meson decays. Ultimately they
also set the scales of nearly all $b$ and $t$\break
 quark branching fractions.  Currently the
uncertainty in
$\calB(\Dz\to\Km\pip)$ -- the best measured of these -- contributes noticeably to the
systematic error in measuring $\Vcb$ in $\BtoDstarlnu$ decays \cite{bdstarlnu}.  We
expect tracking efficiency uncertainties to dominate the systematic errors in
measuring these branching fractions, and tracking uncertainties will be measured using
missing mass techniques.  Ultimately we expect tracking efficiency uncertainties to be 
$\approx 0.2$\% per track.  The \Cleoc\ single and double tag rates for the
reference branching fractions are given in \Tab{tab:dtags}, and the relative errors 
expected are compared to those from the PDG  \cite{pdg} in \Fig{fig:dbferr}.

\section{Measuring $\Vcs$, $\Vcd$, and Form\\ Factors in Semileptonic $D$ Decays}

\begin{figure}[!htb]
\hbox to\hsize{\hss
\includegraphics[width=0.9\hsize]{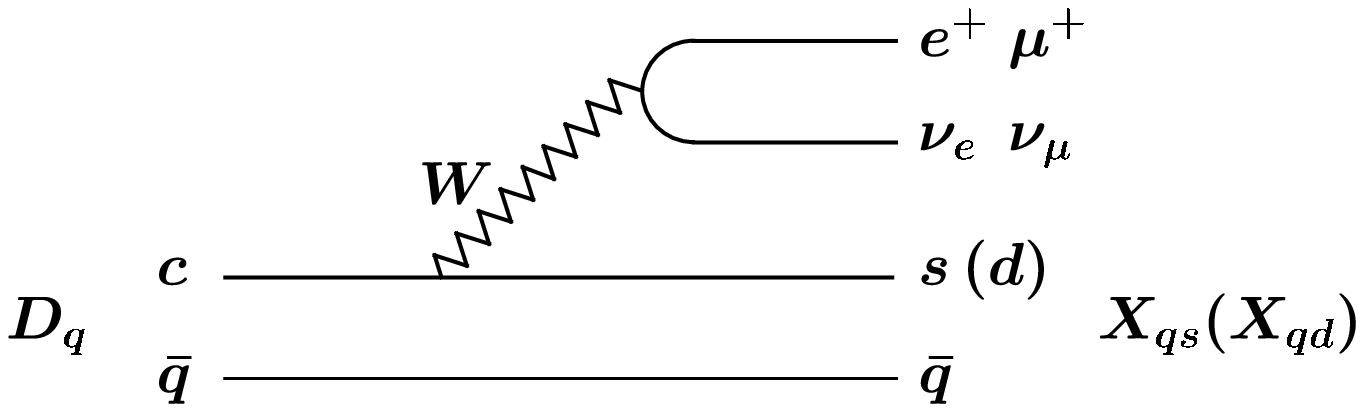}
\hss}
\caption{The Feynman diagram for $D \to X \ell\nu$ decays.}
\label{fig:DXlnu}
\end{figure}

\Fig{fig:DXlnu} illustrates the exclusive semileptonic decays of $\Dz$, $\Dp$, and
$\Ds$ mesons, where $q$ is $u$, $d$, or $s$ for $\Dz$, $\Dp$, or $\Ds$, respectively.
The final state particle $X_{qq'}$ will be $X_{qs}$ for Cabibbo favored $c \to sW$
decays and $X_{qd}$ for Cabibbo suppressed $c \to dW$ decays, with CKM matrix elements
$\Vcs$ and $\Vcd$, respectively, in the decay amplitude.

Exclusive decays depend on the mass-squared ($q^2$) of the virtual $W$ through form
factors {$f(q^2)$}. Decay to a pseudoscalar meson ($P_{qq'}$) involves only one form
factor, and the differential decay width is given by:
\Begeqn
 {\Gamma(D_q \to P_{qq'} \, \ell^+ \nu_\ell)  \over dq^2} = 
{V_{cq'}^2\; p^3 \over 24 \pi^3}~
{|f_{qq'}(q^2)|^2}
\Endeqn
Decay to a vector meson ($V_{qq'}$) involves 3 form factors and a rather more
complicated expression involving 3 decay angles (or 3 other variables) in addition to
$q^2$.  

All of these form factors are nonperturbative QCD functions, whose $q^2$
dependence can be measured but whose normalization or absolute value at some point,
\eg, $q^2_{\rm max}$ must be determined from theory.   
We expect that LQCD will be able to calculate the normalizations of the
form factors $f_{qq'}(q^2)$ with precisions of $\calO(1\%)$.  LQCD should also be able
to predict the $q^2$ dependences of the form factors, so measurements of the $q^2$
dependences can be used to establish the validity of the LQCD calculations.  

\begin{figure*}[!tb]
\hbox to\hsize{\hss
\hfill\includegraphics[width=0.38\hsize]{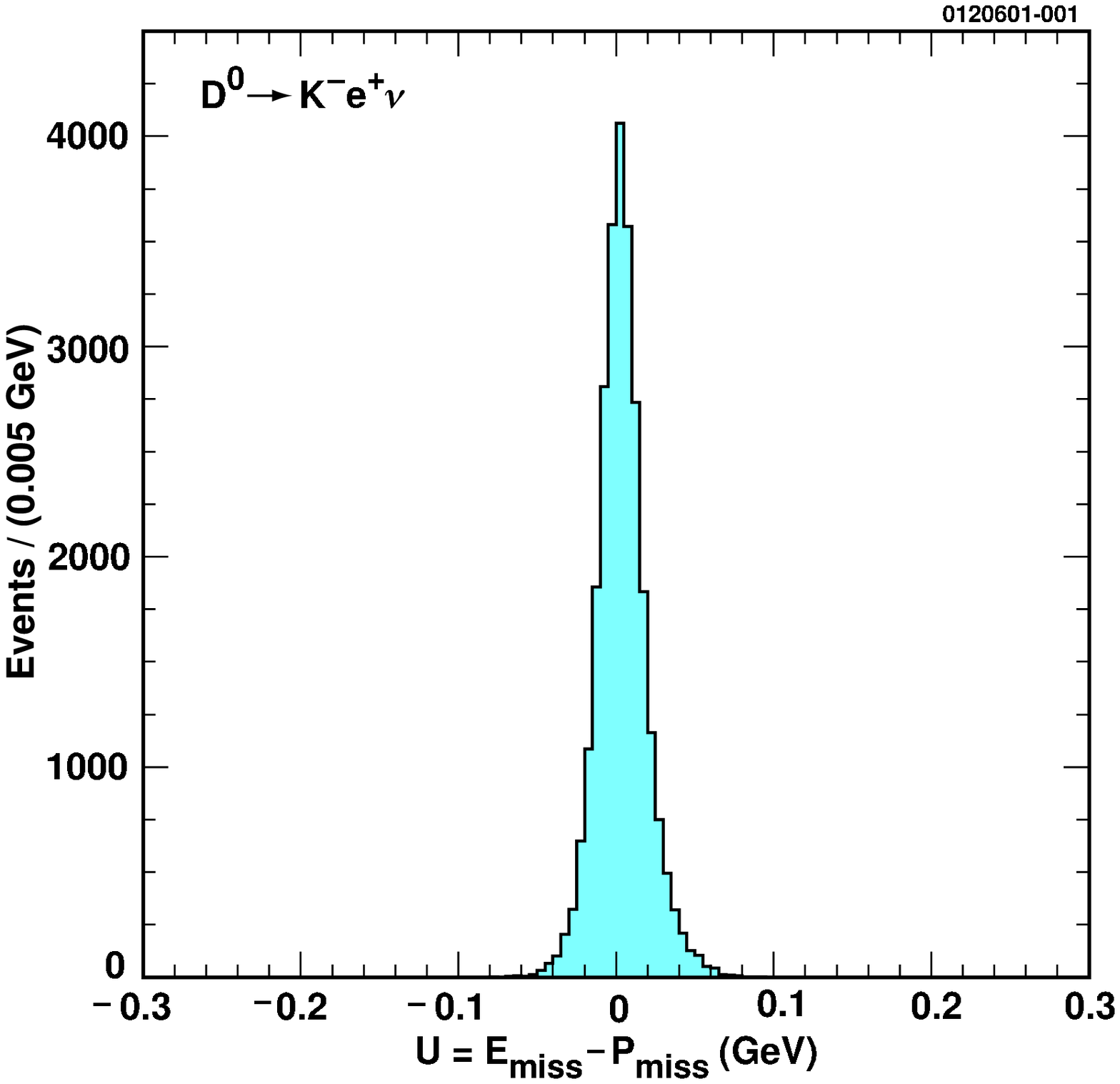}\hfill
      \includegraphics[width=0.38\hsize]{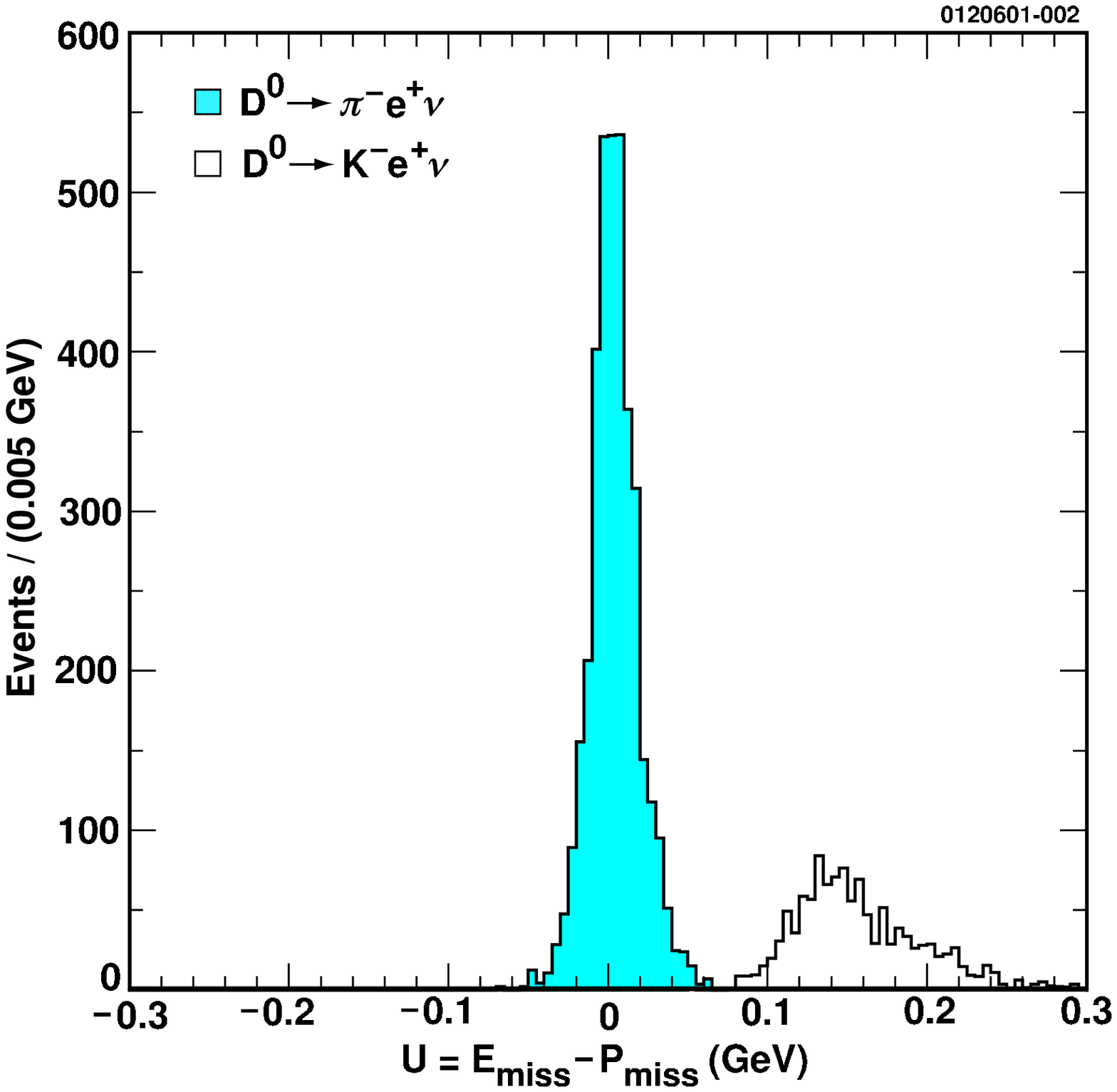}\hfill
\hss}
\caption{Plots of the $U$ distributions for (left) $\Dz \to \Km\ep\nue$ and
(right) $\Dz \to \pim\ep\nue$ decays from Monte Carlo samples corresponding to 1
\fbinv\ of data.  Note the clean separation of $\Dz \to \Km\ep\nue$ background from the
Cabibbo suppressed $\Dz \to \pim\ep\nue$ decay sample.}
\label{fig:udists}
\end{figure*}

\begin{figure*}[!bt]
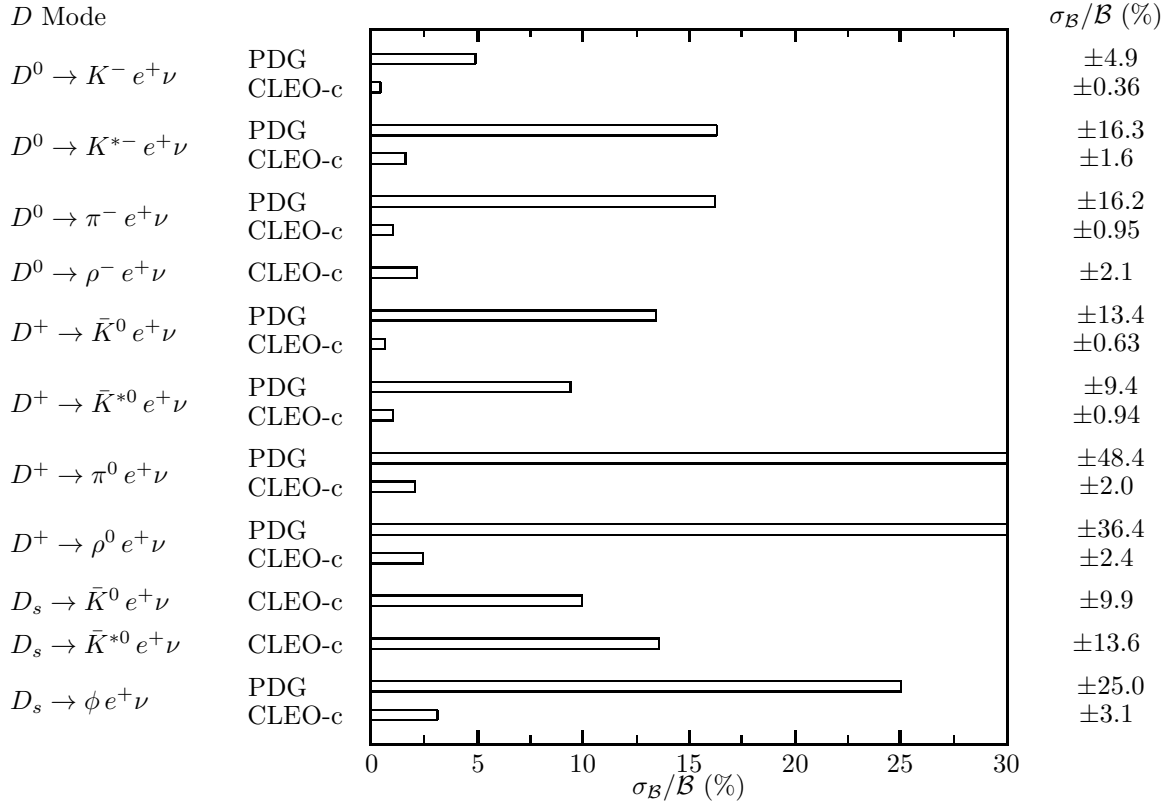

\vspace*{-2ex}
\begin{center}
\setlength{\unitlength}{0.6pt}
\newsummarypar{		400	}{215}{0}{0}{0}{18}																				
\begsummary																																																															
\titline{$D$ Mode}{$\sigma_{\cal B}/{\cal B}$ (\%)}																							
\titskip																							
\datline{	\hfill  PDG~~~~		}{ $\pm 			4.9	$} \lowerlimbar{	65	}{	0	}{	0	} 
\datback																							
\datline{$D^0 \to K^-\, e^+\nu$}{ }																							
\datback																							
\datline{	\hfill    CLEO-c		}{{$\pm 			0.36	$}}  \lowerlimbar{5}{	0	}{	0	}  %
\datskip
\datline{	\hfill    PDG~~~~		}{ $\pm 			16.3	$} \lowerlimbar{	217	}{	0	}{	0	} 
\datback																							
\datline{$D^0 \to K^{*-}\, e^+ \nu$}{ }																							
\datback																							
\datline{	\hfill    CLEO-c		}{{$\pm 			1.6	$}} \lowerlimbar{	21	}{	0	}{	0	} %
\datskip
\datline{	\hfill   PDG~~~~		}{ $\pm 			16.2	$} \lowerlimbar{	216	}{	0	}{	0	} 
\datback																							
\datline{$D^0 \to \pi^-\, e^+\nu$}{ }																							
\datback																							
\datline{	\hfill   CLEO-c		}{{$\pm 			0.95	$}} \lowerlimbar{	13	}{	0	}{	0	} 
\datskip
\datline{$D^0 \to \rho^-\, e^+ \nu$ \hfill   CLEO-c			}{{$\pm 			2.1	$}} \lowerlimbar{	28	}{	0	}{	0	} %
\datskip
\datline{	\hfill   PDG~~~~		}{ $\pm 			13.4	$} \lowerlimbar{	179	}{	0	}{	0	} 
\datback																							
\datline{$D^+ \to \bar{K}^0\, e^+\nu$}{ }																							
\datback																							
\datline{	\hfill   CLEO-c		}{{$\pm 			0.63	$}} \lowerlimbar{	8	}{	0	}{	0	} 
\datskip																							
\datline{	\hfill   PDG~~~~		}{ $\pm 			9.4	$} \lowerlimbar{	125	}{	0	}{	0	} 
\datback																							
\datline{$D^+ \to \bar{K}^{*0}\, e^+\nu$}{ }																							
\datback																							
\datline{	\hfill   CLEO-c		}{{$\pm 			0.94	$}} \lowerlimbar{	13	}{	0	}{	0	} 
\datskip																							
\datline{	\hfill   PDG~~~~		}{ $\pm 			48.4	$} \lowerlimbar{	645	}{	0	}{	0	} 
\datback																							
\datline{$D^+ \to \pi^0\, e^+\nu$}{ }																							
\datback																							
\datline{	\hfill   CLEO-c		}{{$\pm 			2.0	$}} \lowerlimbar{	27	}{	0	}{	0	} 
\datskip																							
\datline{	\hfill   PDG~~~~		}{ $\pm 			36.4	$} \lowerlimbar{	485	}{	0	}{	0	} 
\datback																							
\datline{$D^+ \to \rho^0\, e^+\nu$}{ }																							
\datback																							
\datline{	\hfill   CLEO-c		}{{$\pm 			2.4	$}} \lowerlimbar{	32	}{	0	}{	0	} 
\datskip
\datline{$D_s \to \bar{K}^0 \, e^+ \nu$ \hfill   CLEO-c			}{{$\pm 			9.9	$}} \lowerlimbar{	132	}{	0	}{	0	} %
\datskip																							
\datline{$D_s \to \bar{K}^{*0} \, e^+ \nu$ \hfill   CLEO-c			}{{$\pm 			13.6	$}} \lowerlimbar{	181	}{	0	}{	0	} %
\datskip																							
\datline{	\hfill   PDG~~~~		}{ $\pm 			25.0	$} \lowerlimbar{	333	}{	0	}{	0	} 
\datback																							
\datline{$D_s \to \phi\, e^+\nu$}{ }																							
\datback																							
\datline{	\hfill   CLEO-c		}{{$\pm 			3.1	$}} \lowerlimbar{	41	}{	0	}{	0	} 
\drawframe{$\sigma_{\cal B}/{\cal B}$ (\%)}																							
\percenticks																							
\endsummary
\end{center}\vspace*{-1.5ex}																						
\caption{Relative errors in the $D$ meson semileptonic branching fractions.  PDG
errors are from PDG 2003 and CLEO-c errors are errors expected from CLEO-c.}
\label{fig:dslerrors}
\vspace*{-3ex}\end{figure*}

In \Cleoc\ we can detect semileptonic decays in events with a single hadronic tag
and an $e^\pm$ accompanied by a hadron $X_{qq'}$ or daughters of its
decay.  The branching fractions and $q^2$ dependencies can be measured quite
accurately because: rates are high due to high single tag rates and large 
$D_q \to X_{qq'}\, \ell^+ \nu_\ell$ branching fractions, and background rejection
from kinematics and particle identification is excellent. We find that the variable
$U \equiv E_{\rm miss} - p_{\rm miss}$ (where $E_{\rm miss}$ and $p_{\rm miss}$ are the
missing energy and momentum, respectively) can separate signal from background very
efficiently.  This is illustrated in  \Fig{fig:udists} from a Monte Carlo simulation.  
Note that even the Cabibbo suppressed decay
$\Dz\to\pim e^+\nu$ is separated cleanly from the allowed decay 
$\Dz \to \Km e^+\nu$, whose branching fraction is an order of magnitude larger.
\Fig{fig:dslerrors} illustrates the relative errors expected for a large number of
exclusive $D$ meson semileptonic branching fractions and compares these predictions
to values found in the current PDG summary  \cite{pdg}.

We expect to be able to measure  semileptonic branching fractions  with errors 
$\delta\calB/\calB~ \ltsim~ 1$\% and the exponential slopes ($\alpha$) of form with
errors $\delta\alpha/\alpha \approx 4$\%. These measurements will challenge LQCD
theorists to calculate form factors with precisions of
$\calO(1\%)$.  If the challenges are met, \Cleoc\ measurements of semileptonic $D$
branching fractions will provide values of $\Vcs$ and $\Vcd$ with errors 
$\ltsim~2$\%.   \Tab{tab:vcqserrs} shows the contributions of experimental
uncertainties to the uncertainties in measuring $\Vcs$ and $\Vcd$ and compares these
uncertainties with those from unitarity. Consistency of $\Vcs$ and $\Vcd$ results from
many $\Dz$ and $\Dp$ modes and with unitarity will help to verify
experimental systematic errors and LQCD calculations of form factors.

\vspace*{-1ex}

\section{Measuring $f_{\Dp}$ and $f_{\Ds}$ in Leptonic $D$ Decays}

\vspace*{-1ex}

\begin{figure*}[!htb]
\hbox to\hsize{\hss
\hfill\includegraphics[width=0.4\hsize]{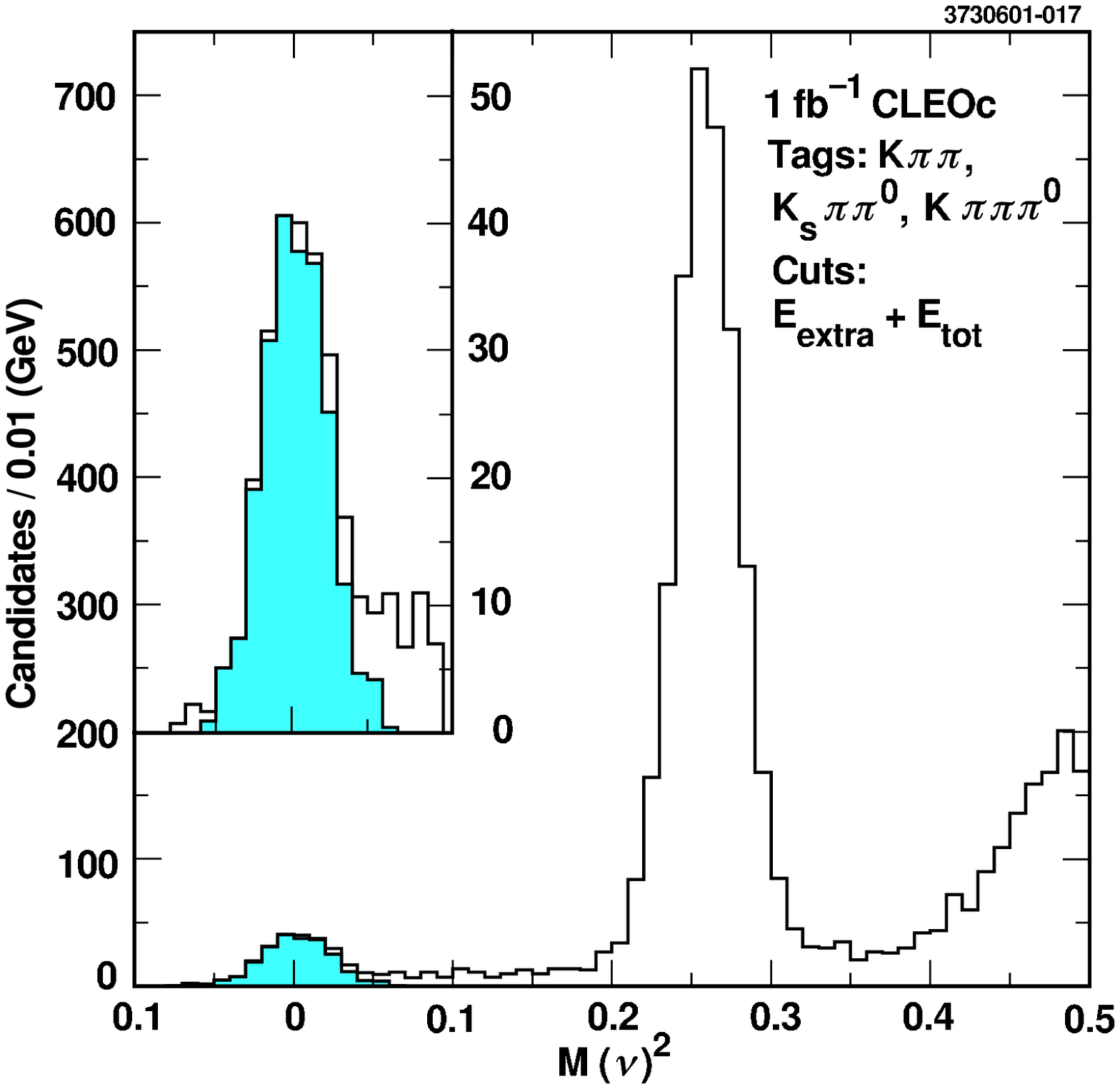}\hfill
      \includegraphics[width=0.38\hsize]{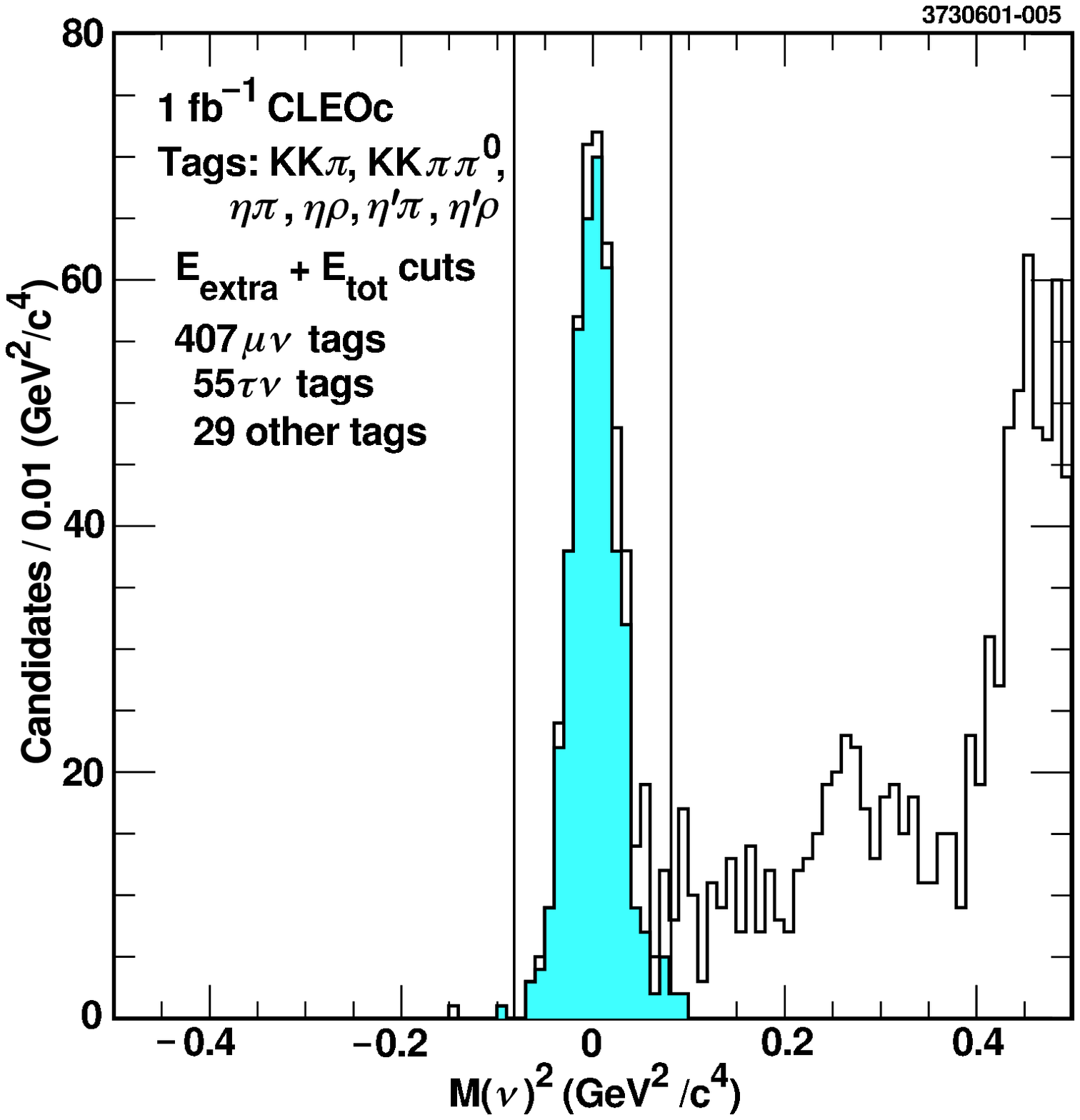}\hfill
\hss}
\caption{Plots of the $M(\nu)^2$ distributions for (left) $\Dp \to \mup\numu$ and
(right) $\Ds \to \mup\numu$ decays.  These Monte Carlo simulations correspond to 1
\fbinv\ of data, not the expected 3 \fbinv.}
\label{fig:Dmnusq}
\vspace*{-2ex}\end{figure*}

\Fig{fig:Bptolnu} also illustrates the Feynman diagram for $\Dqp \to \ell^+ \nu_\ell$
decay -- where $D_q$ is either $\Dp$ or $\Dsp$ -- if $\Bp$ is replaced with $\Dqp$.  The
factor
$f_{D_q}V_{cq}$ occurs in the decay amplitude for the $c\bar{q}W$ vertex, and decay
widths for leptonic
$D^+$ and $D_s^+$ decays are given by \Eqn{eq:fb} with $M_B$ replaced by $M_{\Dq}$ and
$\fB\Vub$ replaced with $\fDq\Vcq$.
Therefore, measurements of
$\calB(D^+ \to \ell^+ \nu_\ell)$ and $\calB(D_s^+ \to \ell^+ \nu_\ell)$ can be used
to determine $f_{D^+} \Vcd$ and $f_{D_s} \Vcs$, respectively.  The branching fractions
for 
$\Dqp \to \ell^+ \nu_\ell$ are much larger than the branching fractions for 
$B^+ \to \ell^+ \nu_\ell$, because $V_{cq}$ is much larger than $\Vub$.  Using 
reasonable guesses for $f_{D^+}$ and $f_{D_s}$ (220 MeV and 260 MeV, respectively)
we estimate\\
$\calB(D^+ \to \mu^+ \nu_\mu) \sim 4 \times 10^{-4}$ and\\ 
$\calB(\Ds^+ \to \mu^+ \nu_\mu) \sim 6 \times 10^{-3}$.\\   
These branching fractions, the high rates of $D\Dbar$ production, and high tagging
efficiencies all combine to enable precision measurements of 
$f_{D_q} \Vcq$.  \Fig{fig:Dmnusq} illustrates the separation of 
$\Dqp \to \mu^+ \nu_\mu$ decays from background that can be achieved with tagged
samples of $D_q$ decays accompanied by a single $\mu$.  The mass $M(\nu)$ of the
missing $\nu$ is computed from the beam energy, the momentum of the primary tag, and
the momentum and energy of the observed $\mu$.
     
Since we will measure $\Vcd$ and
$\Vcs$ accurately with semileptonic $D$ decays, and have unitarity of the CKM matrix
to check these values, we can determine $f_{D^+}$  and $f_{D_s}$  with errors
$\calO(1\%)$. \Tab{tab:fDqerrs} shows that $f_{D^+}$ and $f_{D_s}$ can be measured
with precisions $\sim 2\%$ with the full \Cleoc\ data samples when all sources of
uncertainty are taken into account.  These measurements will challenge LQCD theorists
to compute $f_{D^+}$ and $f_{D_s}$ with uncertainties $\sim 2\%$ and lead to an
understanding of the level of reliability of corresponding LQCD $\fB$ calculations. 
Furthermore, LQCD calculations of the ratio $\fB/\fDp$ are expected to be more
reliable than calculations of either $\fB$ or $\fDp$ \cite{fB/fD}, so precision
measurement of $\fDp$ can be used to derive an accurate value of
$\fB$.  

\begin{table*}[!htb]
\caption{Contributions to errors in $\Vcs$ and $\Vcd$ expected from 3~\fbinv\ of
$\psi(3770)$ -- $\Dz\Dzbar$ and $\Dp\Dm$ -- \Cleoc\ data.  In this table, $\calB$,
$\tau$, and $\epsilon$ are the relevant branching fractions, lifetimes, and detection
efficiencies, respectively.}
\label{tab:vcqserrs}
\begin{center}
\Begtabular{ccccccc}
Decay Mode & ~~~$V$~~~ & ~~$\frac{1}{2}(\delta{\cal{B}}/{\cal{B}})$~~ & 
 ~~$\frac{1}{2}(\delta \tau/\tau)$~~ & ~~$\frac{1}{2}(\delta\epsilon/\epsilon)$~~ & ~~$\delta V/V$~~ &
Unitarity \\ \hline
$D^0\to \Km  e^+ \nu$    & $\Vcs$ &  0.2\% & 0.35\% & 0.45\% & {0.6\%} & { 0.1\%} \\ 
$D^+\to \Kzbar  e^+ \nu$ & $\Vcs$ &  0.3\% & 0.6\%  & 0.45\% & {0.8\%} & { 0.1\%} \\ \hline
$D^0\to \pim e^+ \nu$    & $\Vcd$ &  0.5\% & 0.35\% & 0.45\% & {0.8\%} & { 1.1\%} \\
$D^+\to \piz e^+ \nu$    & $\Vcd$ &  1.0\% & 0.6\%  & 0.45\% & {1.3\%} & { 1.1\%} \\ 
\Endtabular
\end{center}
\vspace*{-2ex}\end{table*}

\begin{table*}[!tb]
\caption{Contributions of the major uncertainties to errors in $\fDp$ and $\fDs$
expected from 3~\fbinv\ each of $\Dp\Dm$ and $\Dsp\Dsm$ \Cleoc\ data.  In this
table, $\calB$ and $\tau$ are the relevant branching fractions measured in \Cleoc\
and the lifetimes, respectively.  The last column gives the uncertainties from the
current PDG summary.}
\label{tab:fDqerrs}
\begin{center}
\Begtabular{ccccccccc}
Decay Mode & Signal & Bkg  & ~$\frac{1}{2}(\delta{\cal{B}}/{\cal{B}})$~
& ~$\frac{1}{2}(\delta \tau/\tau)$~ & ~$\delta|V_{cq}|/|V_{cq}|$~ 
& ~$\delta f_{D_q}/f_{D_q}$~ & & PDG \\ \hline
$D^+\to\mu^+\nu$ & ~~~672 & ~90  & 1.9\% & 0.6\% & 1.1\% & {2.3\%} & $f_{\Dp}$ & { ---} \\
\hline
$D_s^+\to\mu^+\nu$ & 1,221 & 252  & 1.4\% & 1.0\% & 0.1\% & {1.7\%} & $f_{D_s}$ & { 35\%} \\ 
$D_s^+\to\tau^+\nu$ & 1,740 & 114 & 1.2\% & 1.0\% & 0.1\% & {1.6\%} & $f_{D_s}$ & { 60\%} \\
\Endtabular
\end{center}
\vspace*{-2ex}\end{table*}

\section{CKM Element Uncertainties}

\begin{figure}[!htb]
\hbox to\hsize{\hss
\includegraphics[width=\hsize]{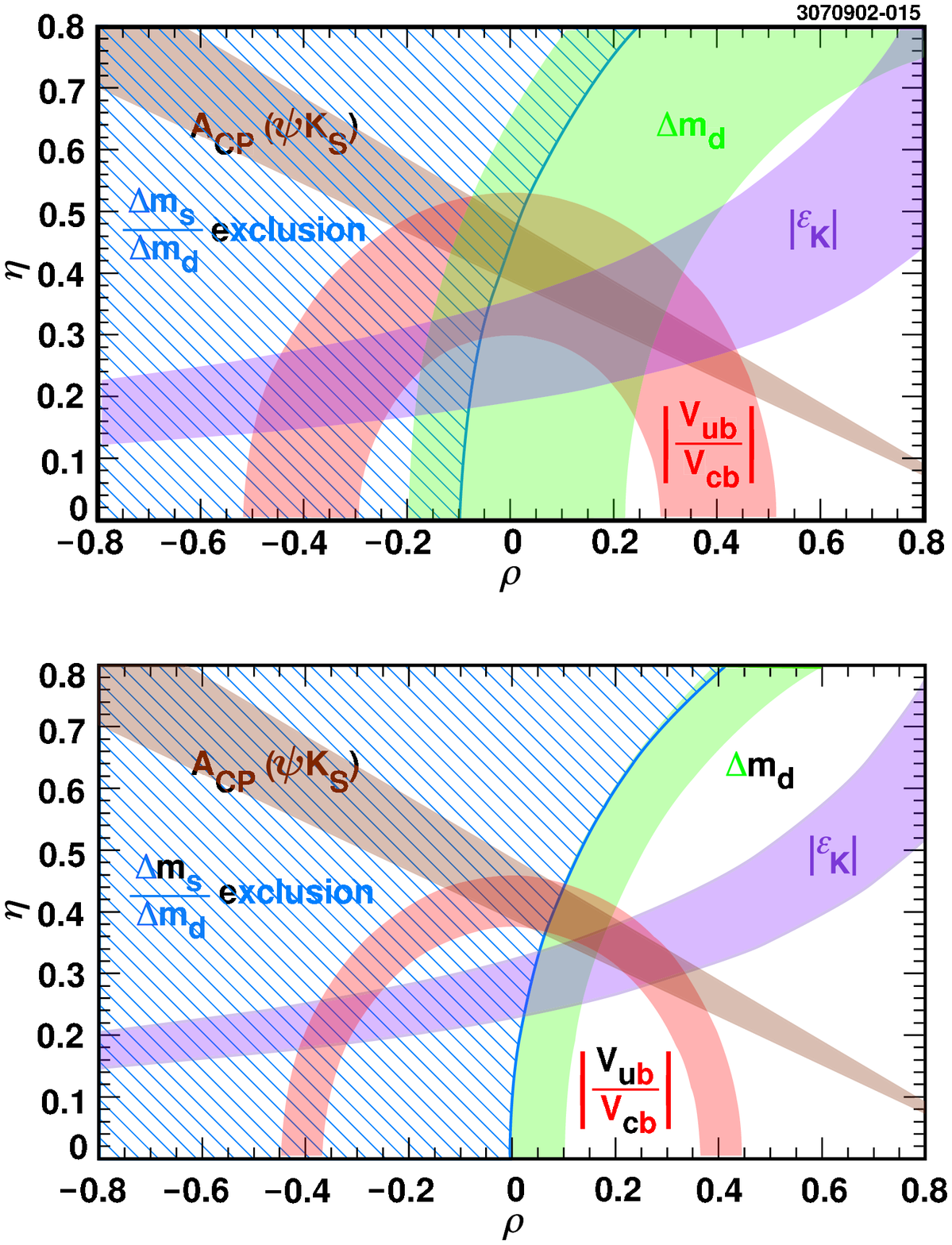}
\hss}
\caption{Plots of allowed regions in the  $\rho$-$\eta$ using current experimental
measurements and (top) quite conservative current theoretical uncertainties and
(bottom) theoretical uncertainties resulting from the \Cleoc\ program and improved
LQCD calculations.  Note that the improvement in the
$\varepsilon_K$ band will not result directly from the effect of \Cleoc\ measurements on
LQCD, but from more general progress in LQCD.}
\label{fig:ckmplots}
\end{figure}

\Fig{fig:ckmplots} illustrates the present uncertainties in CKM matrix elements
plotted in the $\rho$-$\eta$ plane \cite{wolfenstein} and the uncertainties that could
result from the verification of LQCD calculations by \Cleoc.  The top plot uses
current experimental uncertainties and quite conservative current theoretical
uncertainties.  An up-to-date overview of uncertainties of CKM parameters in the
$\rho$-$\eta$ plane can be found in these proceedings \cite{lubicz}.  The bottom plot use
the same experimental uncertainties but  theoretical uncertainties of
$\calO(1\%)$; in particular, uncertainties of 2\% for decay constants and bag parameters,
and 3\% for semileptonic form factors.  It is clear that the \Cleoc\ program can have a
substantial impact on our understanding of the CKM matrix.

Other experimental programs can also contribute significantly to precision measurements
of CKM matrix elements:
\Begitem

\item BEPCII and BESIII\\
BEPCII will be a new $e^+e^-$ collider in Beijing operating in the charm threshold
region with anticipated luminosity at least three times that of \Cesrc. Many
capabilities of the proposed\break BESIII detector are comparable to that of the
\hbox{\Cleoc} detector. The Chinese government approved the BEPCII proposal in
February and the Beijing group expects to turn on in about 5 years.

\item BaBar, BELLE, and FOCUS\\
Absolute $D$ branching fractions are hard to measure in the $\Upsilon$ region or at
Fermilab.  However, precision measurements of the ratios of branching fractions and
the  $q^2$ dependence of form factors in semileptonic decays can constrain \Cleoc\ and
BESIII results.   Improved measurements of charm lifetimes will also be important if
BESIII is able to reduce systematic errors substantially below those of \Cleoc. 
\Enditem

\vspace*{-1ex}

\section{Summary and Outlook}

\vspace*{-1ex}

Nonperturbative QCD parameters are needed to extract
$\Vcb$, $\Vub$, and $\Vtd$ from $B$ physics measurements.
Even with CLEO's luminosity, residual theoretical uncertainties are already
comparable to experimental errors.
Substantial theoretical progress will be required in order to fully
benefit from the large luminosities being accumulated by BaBar and Belle.
In the $D$ meson sector, \Cleoc\ can measure absolute branching fractions,
semileptonic decays, and leptonic decays ($D \to \bar\ell \nu$) with
$\sim 1$\% precision.

This program can motivate Lattice QCD theorists to attempt to reach comparable precision
in calculating the nonperturbative QCD parameters involved in these $D$ decay
measurements -- particularly semileptonic decay form factors and $f_{D_{(s)}}$. Success
in this program will build confidence in applying these calculations to the $B$ sector
for measurements of $\Vcb$, $\Vud$, and $\Vtd$.
\Cleoc\ and (later) BESIII will likely dominate absolute branching fractions 
measurements.  FOCUS, BaBar, and BELLE have contributed or will contribute significantly
to charm lifetimes, relative branching fractions and form factor measurements, and
searches for new physics.

As the \Cleoc\ and LQCD programs gain momentum, we can expect very fruitful
interactions between theory and experiment leading to substantial improvements in our
knowledge of CKM matrix elements.

\vspace*{-1ex}

\section*{Acknowledgements}

\vspace*{-1ex}

I am delighted to acknowledge the many contributions of my CLEO and
CESR colleagues which led to the results and ideas expressed in this contribution.  I
appreciate the support of the National Science Foundation for the CESR/CLEO
program and NSF and U.S. Department of Energy for support of my CLEO collaborators. 
Finally, I want to thank the organizers and their staff who have worked so hard to
make this workshop pleasant and successful. 

\def\Journal#1#2#3#4{{#1} {\bf #2}, #3 (#4)}
\def\Report#1#2#3{#1 Report No.\ #2 (#3)}
\def\Journalmore#1#2#3{{\bf #1}, #2 (#3)}

\def\CLEO{CLEO Collaboration}
\def\EPJC{{\it E.\ Phys J.} C}
\def\NCA{\it Nuovo Cimento}
\def\NIM{\it Nucl.\ Instrum.\ Methods}
\def\NIMA{{\it Nucl.\ Instrum.\ Methods} A}
\def\NPB{{\it Nucl.\ Phys.} B}
\def\PLB{{\it Phys.\ Lett.} B}
\def\PRL{\it Phys.\ Rev.\ Lett.\ }
\def\PRD{{\it Phys.\  Rev.} D}
\def\ZPC{{\it Z.\ Phys.} C}


\vspace*{-1ex}

\begin{thebibliography}{99}
\bibitem{hplqcd}C.T.H. Davies \etal, hep-lat/0304004.
\bibitem{cleocyb}\Cleoc\ Collaboration, R.A. Briere \etal, {\it CLEO-c and CESR-c: A New
Frontier of the Weak and Strong Interactions}, Cornell Report CLNS 01/1742, Revised
October 2001, available from a link at
{\tt http://www.lns.cornell.edu/}
\bibitem{ckmyb}M.~Battaglia {\it et al.},
{\it The CKM matrix and the Unitarity Triangle},
{arXiv:hep-ph/0304132}. 
\bibitem{pdg}Particle Data Group, K. Hagiwara \etal, \Journal{\PRD}{66}{010001}{2002}.
\bibitem{dgcmoments}D.G. Cassel, these proceedings.
\bibitem{lkg}L.K. Gibbons, these proceedings.
\bibitem{bdstarlnu}CLEO Collaboration, R.A. Briere \etal, 
\Journal{\PRL}{89}{081803}{2002} and N.E. Adam \etal, \Journal{\PRD}{67}{032001}{2003}.
\bibitem{bulnu}CLEO Collaboration, S.B. Athar \etal, 
\Report{Cornell}{CLNS 03/1819, CLEO 03-05}{2003} submitted to \PRD.
\bibitem{ichep:tomasz}T. Skwarnicki, Proceedings of ICHEP 2002. 
\bibitem{markiii}MARK III Collaboration, R.M. Baltrusaitis \etal,
\Journal{\PRL}{56}{3140}{1986}.
\bibitem{fB/fD}D. Becirevic and P. MacKenzie, these proceedings.
\bibitem{wolfenstein}L. Wolfenstein, \Journal{\PRL}{51}{1945}{1983}.
\bibitem{lubicz}V. Lubicz, these proceedings.
\end{thebibliography}
\end{document}